\documentclass[twocolumn,showpacs,reprint,amsmath,amssymb,aps,prl]{revtex4}
\usepackage{graphicx}% Include figure files
\usepackage{dcolumn}% Align table columns on decimal point
\usepackage{bm}% bold math
\usepackage{subfigure}
\usepackage{natbib}
\usepackage{multirow}
\usepackage{soul}
\usepackage[usenames,dvipsnames]{xcolor}
\usepackage{soul}
\usepackage{newfloat}

\DeclareFloatingEnvironment[name={Fig. S}]{suppfigure}

\begin{document}
\title{\center{Hydrodynamic heat transport regime in bismuth~: a theoretical viewpoint}}
\author{Maxime Markov$^1$}
\email{maksim.markov@polytechnique.edu}
\author{Jelena Sjakste$^1$}
\author{Giuliana Barbarino$^1$}
\author{Giorgia Fugallo$^2$}
\author{Lorenzo Paulatto$^3$}
\author{Michele Lazzeri$^3$}
\author{Francesco Mauri$^4$}
\author{Nathalie Vast$^1$}
\affiliation{\vskip 0.2cm
$^{1}$ \'Ecole Polytechnique, Laboratoire des Solides Irradi\'{e}s, CNRS UMR 7642, CEA-DSM-IRAMIS, Universit\'{e} Paris-Saclay, F91128 Palaiseau c\'edex, France,}
\vskip 0.05cm
\affiliation{$^{2}$ CNRS, LTN UMR 6607, PolytechNantes, Universit\'{e} de Nantes, Rue Christian Pauc, 44306 Nantes c\'edex 3, France}
\vskip 0.05cm
\affiliation{$^{3}$ Sorbonne Universit\'es, UPMC Univ. Paris 06, CNRS UMR 7590, MNHN, IRD UMR 206 Institut de Min\'eralogie, de Physique des Mat\'eriaux et de Cosmochimie, 75005 Paris, France}
%\affiliation{$^{2}$ Sorbonne Universit\'es, UPMC Univ. Paris 06, CNRS UMR 7590, MNHN, IRD, IMPMC, 75005 Paris, France}
\vskip 0.05cm
\affiliation{$^{4}$ Dipartimento di Fisica, Universit\`a di Roma La Sapienza, Piazzale Aldo Moro 5, I-00185 Roma, Italy.}
\date{\today}

\begin{abstract}
Bismuth is one of the rare materials in which second sound has been experimentally observed.  
Our \textit{exact} calculations of thermal transport with the Boltzmann equation 
predict 
the occurrence of this Poiseuille phonon flow between $\approx$~1.5~K~and~$\approx$~3.5~K, in sample size of 3.86~mm and 9.06~mm, 
in  
consistency with the experimental observations.
Hydrodynamic heat flow characteristics are given for any temperature~: heat wave propagation length, drift velocity, Knudsen number. 
We discuss a Gedanken experiment allowing to assess the presence of a hydrodynamic regime in any bulk material. 
\end{abstract}

%\pacs{L6-59: Thermal and Nonelectronic Transport in Condensed Matter}

\maketitle

Currently a lot of attention is devoted to the study of phonon-based heat transport regimes in 
nanostructures~\cite{Cahill:2014,Volz:2016,Chang:2008,Yang:2010}.
Of particular interest is the hydrodynamic regime in which a number of fascinating phenomena such as Poiseuille's phonon flow and second sound occur, 
and where temperature fluctuations are predicted to propagate 
as a true temperature wave of the form  $e^{i (\mathbf{k} \cdot \mathbf{r} - \omega t)}$~\cite{Guo:2015}. 
The theoretical study of the hydrodynamic regime 
has encountered a renewed interest in graphene nanoribbons, 
where the breakdown of the diffusive Fourier law in favor of the second sound propagation has been predicted~\cite{Zhang:2011d,Fugallo:2014,Lee:2015,Cepellotti:2015}. 
Bismuth has the particularity to be a semimetal with relatively low carrier concentrations so that the dominant mechanism for heat conduction
at low temperatures is \textit{via}  phonons~\cite{Narayanamurti:1972,Issi:1979}.
Together with solid helium~\cite{Ackerman:1966} and NaF~\cite{Jackson:1970}, it is one of the rare 
materials that are sufficiently isotopically pure so that  second sound 
could be observed. The degree of physical and chemical perfection that has been 
achieved in Bi crystals is so high that 
also 
\textit{transitions} between the various regimes have been experimentally observed with the increase of the (yet cryogenic) temperature~:  
from heat transport  
\textit{via} ballistic phonons,  to the regime of Poiseuille's flow with 
second sound, 
to the diffusive (Fourier) propagation~\cite{Narayanamurti:1972}.  

Neither the conditions for the occurrence of the hydrodynamic regime nor the transition temperatures have ever been supported  
by a theoretical work in one of the above-cited 3-D materials. 
So far, 
phonon hydrodynamics 
has been studied with the lattice Boltzmann formalism for a model dielectric material 
with an \textit{ad hoc} three-phonon collision term and no resistive processes~\cite{Guyer:1994}. The transition to the kinetic regime 
has been modeled 
in group IV semiconductors 
through a hydrodynamic-to-kinetic switching factor proportional to the ratio of normal and resistive scattering rates~\cite{DeTomas:2014,DeTomas:2014b,DeTomas:2015}. 
A 
review on 
advances in phonon hydrodynamics points out the lack of a widely applicable 
hydrodynamic model which would consider all of the normal and resistive processes~\cite{Guo:2015}.

In this work, a  major advance consists in accounting for the phonon repopulation by the normal processes in the framework of the exact variational solution of the Boltzmann transport equation (V-BTE)~\cite{Omini:1995,Fugallo:2013}, coupled to the \textit{ab initio} description of anharmonicity~: three-phonon collisions turn
 out to be particularly strong at low temperatures, and lead to the creation of new phonons in the direction of the heat flow (normal processes) which enhance the heat transport. This induces time- and length-scales over which heat carriers behave collectively and form a hydrodynamic flow that cannot be described by independent phonons 
with their own energy and lifetime. In other words the single mode approximation (SMA), valid for the phonon gas model, breaks down. 
The resistive processes are entirely controlled by few phonon-phonon anharmonic processes 
which 
lead to the creation of phonons in the direction opposite to the heat flow (Umklapp processes),
and by extrinsic processes coming from phonon scattering by the sample
boundaries. 

The characterization of heat transport regimes, and in particular of the transition between the hydrodynamic and kinetic regimes, 
is the main focus of present work.  
We discuss several methods to define the hydrodynamic regime, and provide the link with macroscopic scale quantities~\cite{Guo:2015} 
like  
Knudsen number and drift velocity. 
In particular, we extract  the heat wave propagation length (HWPL) directly from the lattice thermal conductivity (LTC)  calculated with V-BTE. 
We argue that
our method to extract
the HWPL from the LTC in
samples of different sizes, combined to a measurement of the average phonon mean free path,
can be viewed as
a Gedanken experiment which could allow to determine
the transition from the hydrodynamic to kinetic regime
in any material.

Several criteria 
are 
used  in order to identify the hydrodynamic to kinetic transition. 
First, the picture of the heat carried by single (uncorrelated) phonons with finite lifetimes, is valid in the 
kinetic regime only.
Thus, a significant difference between the LTC obtained by a 
solution of  V-BTE  and the one obtained in the single mode approximation (SMA-BTE) is the indication that the  
hydrodynamic regime is achieved. 
Second, we compare the thermodynamic averages of the phonon-scattering rates for normal and  resistive processes $\Gamma^{n}$ and  $\Gamma^{U}$, 
and the hydrodynamic regime occurs when~\cite{Guyer:1966}
\begin{equation}
\Gamma^{U}_{av} \ll \Gamma^{n}_{av}. 
\label{eq:condition_hydro}
\end{equation}

Then, we address the question of the occurrence of Poiseuille's flow inside the hydrodynamic regime.  
Here as well, various  methods are employed, which now account for the additional scattering rate by sample boundaries  $\Gamma^{b}$. 
We first use  Guyer's conditions~\cite{Guyer:1966}, 
\begin{equation}
 \Gamma^{U}_{av}   \, \, <  \,  \,  \Gamma^{b}_{av}  \, \, <  \, \,  \Gamma^{n}_{av} \, 
\label{eq:condition_Poiseuille}
\end{equation}
and  find the temperature interval in which
second-sound is calculated to be observable. 
In the second method, we 
extract the heat wave propagation length directly from the LTC calculated with V-BTE 
and compare it to the sample size which
sets the threshold for 
the second sound observability. Above the threshold, the heat-wave is damped before reaching the sample boundary. 

\begin{figure} [t]
\includegraphics[width=0.45\textwidth]{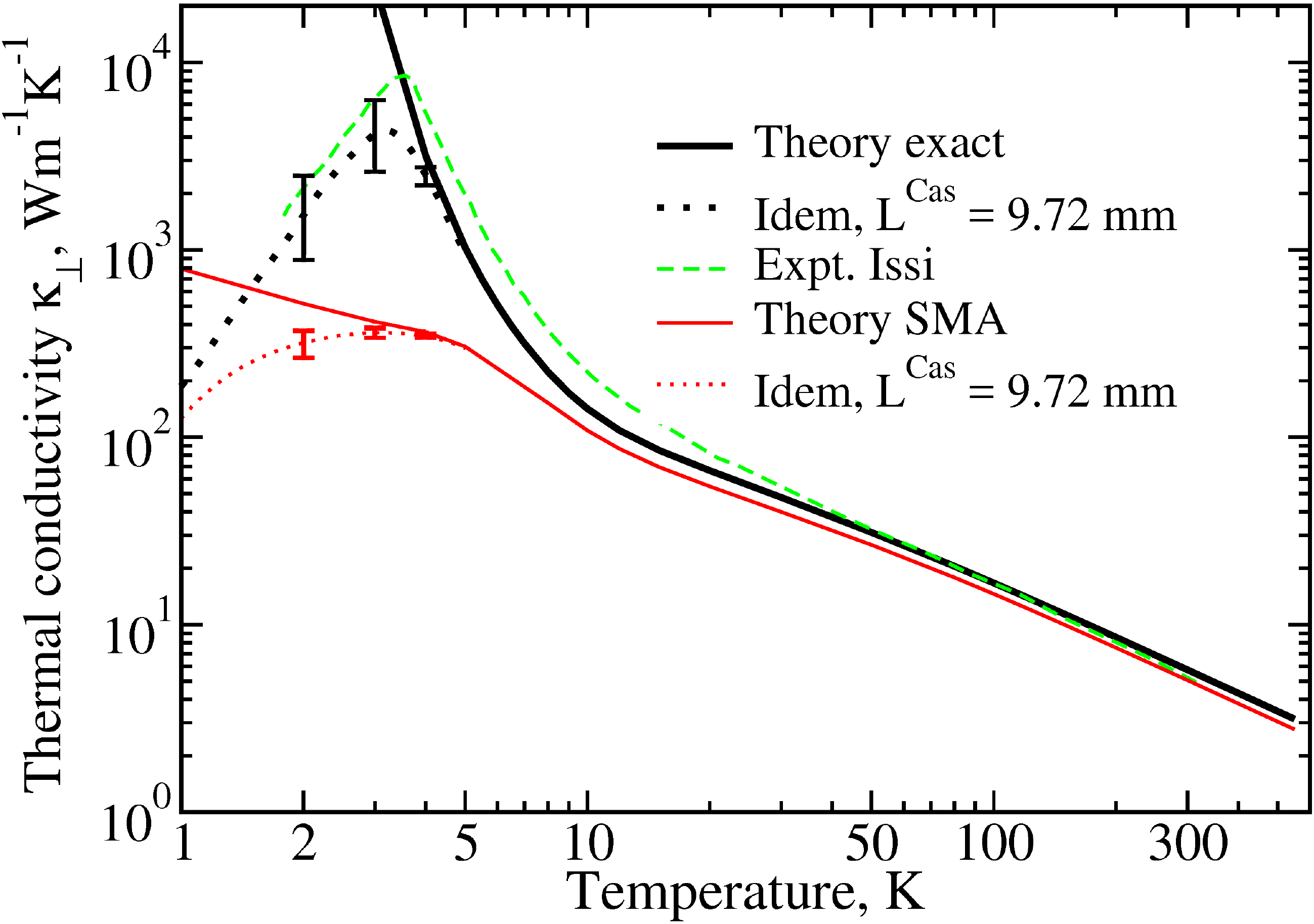}
\caption{\label{fig:thermalk-bulk}
Temperature dependence of the LTC in the
binary direction for a single crystal without (solid lines) or with (dotted lines) millimeter-sized sample boundaries (MSSB).   
Black curves~: exact variational calculation (V-BTE). Red curves~: single mode approximation (SMA-BTE).  
MSSB modeled with the wire geometry and $L^\mathrm{Cas}$~$=$~9.72~mm~\protect\cite{Markov:Note:2015:Lcas_macro_one}.
Green dashed lines~:  LTC extracted by us from expt. of Ref.~\protect\onlinecite{Uher:1974} for $T>20$~K;
for $T<20$~K, LTC from a sample  having a rectangular cross-section $8.8\times8.6$ mm$^{2}$ (Ref.~\protect\onlinecite{Issi:1979}).
We used the T-independent bulk value of 6~W(K.m)$^{-1}$~\protect\cite{Markov:2016,Markov:2016b} of the electronic contribution 
to extract the LTC from the total thermal conductivity of Bi~\protect\cite{Issi:1979}. The error bar in our calculations results 
from the variation of the geometrical factor $\frac{1}{F}$=2$\pm$1.
}
\end{figure}

The thermodynamic averages of phonon scattering rates for normal, Umklapp and boundary collisional processes  that 
condition the transport regime read 
\begin{equation} \label{eq:average_scatt_rate}
\Gamma^{i}_{av} = \frac{\sum_{\nu} C_{\nu}\Gamma_{\nu}^{i}}{\sum_{\nu} C_{\nu}}
\end{equation}
where  $C_{\nu}$ is the specific heat (see below) of the phonon mode $\nu = \{\mathbf{q}j\}$ and the index $i = n, U, b$ stands for normal, Umklapp and extrinsic (boundary) scattering respectively. 
Besides the scattering rate (or inverse relaxation time), the quantities characterizing heat transport are the drift velocity $v$ of the heat carriers 
defined below,  
and the phonon propagation length $\lambda = v \, \Gamma_{av}^{-1}$ 
which is the characteristic distance that 
heat carrying phonons  cover before damping.  As a source of damping, we consider, in infinite samples, either Umklapp processes only 
\begin{equation} \label{eq:propagation_length_hydro}
\lambda_{hydro}(\infty)=v/\Gamma_{av}^{U}, 
\end{equation}
or  their combination with normal processes through Matthiessen's rule 
\begin{equation} \label{eq:propagation_length_gas}
\lambda_{gas}(\infty)=v/(\Gamma_{av}^{U}+\Gamma_{av}^{n})\,.
\end{equation}
When scattering by sample boundaries is accounted for, the phonon propagation length reads  $\lambda (L^\mathrm{Cas})$ instead of $\lambda (\infty)$ 
in eqs.~\ref{eq:propagation_length_hydro}$-$\ref{eq:propagation_length_gas}, 
where Casimir's length $L^\mathrm{Cas}$ represents the smallest dimension of the sample or nanostructure. 

In bismuth the transport is anisotropic and has components along the trigonal axis ($\parallel$) and perpendicular ($\perp$) to it,
i.e. along the binary and bisectrix directions.
The drift velocity in these directions reads~\cite{Cepellotti:2015}
\begin{equation}
v_{j}^2 = \frac{\sum_{\nu} C_{\nu} \, \, \mathbf{c}_{\nu \, j}\cdot \mathbf{c}_{\nu \, j}}{\sum_{\nu}C_{\nu}},
\label{eq:v}
\end{equation}
where $j$ stands for $\parallel$ or $\perp$ direction and $\mathbf{c}_{\nu}$  is the  phonon group velocity. 
In the thermodynamic averages
the specific heat of a phonon mode is calculated as
$C_{\nu}$~=~$n_{\nu}^{0}(n_{\nu}^{0}+1)\frac{(\hbar\omega_{\nu})^2}{k_{B}T^2}$, 
where  $n^{0}$ stands for the temperature (T) dependent Bose-Einstein phonon occupation number and
$\omega_\nu$ is the phonon frequency.

\begin{figure} [t]
\includegraphics[width=0.5\textwidth]{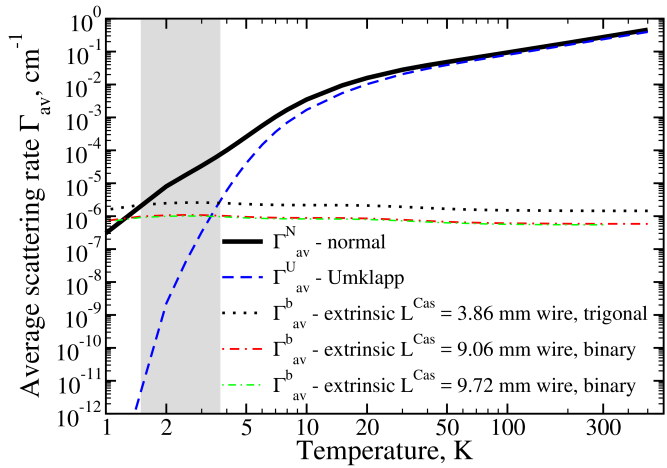}
\caption{\label{fig:lw_avg}
Temperature dependence of the thermodynamic average of the anharmonic scattering rates for  normal and Umklapp processes
(resp. black solid and blue dashed line) and of the (boundary) extrinsic scattering rates (ESR) (dashed black and dot-dashed red and green lines). 
ESR have been calculated for a wire geometry using $L^\mathrm{Cas}$~$=$~$3.86$ mm~\protect\cite{Markov:Note:2016:Lcas_second_sound_1}
and  $L^\mathrm{Cas} = 9.06$ mm~\protect\cite{Markov:Note:2016:Lcas_second_sound_2}, 
and with $L^\mathrm{Cas}$~$=$~$9.72$~mm~\protect\cite{Markov:Note:2015:Lcas_macro_one} as in Fig~\ref{fig:thermalk-bulk}. 
ESR for $L^\mathrm{Cas}$~$=$~$9.06$ and $L^\mathrm{Cas}$~$=$~$9.72$~mm are hardly distinguishable on the scale of the figure. 
The shaded region corresponds to the temperature interval in which a second sound peak has been reported,  1.5~K $< T <$ 3.5~K  for a sample of length 
3.86~mm in the trigonal propagation direction~\cite{Narayanamurti:1972}. 
}
\end{figure}

\begin{figure} [t]
\includegraphics[width=0.5\textwidth]{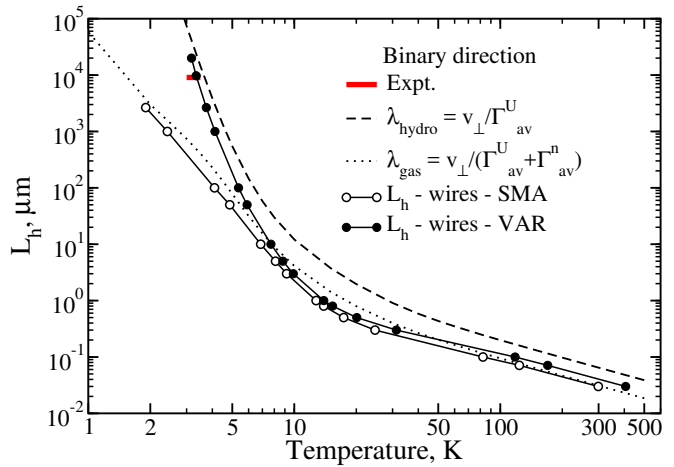}
\caption{\label{fig:regime_Lcas_T}
Heat wave propagation length $L_h$ extracted from the LTC calculations in the binary direction 
as a function of temperature.
Solid line with black filled disks~: 
$L_h$ obtained with V-BTE, accounting for phonon repopulation. Solid line with empty circles : 
$L_h$ obtained with SMA-BTE. 
Black dashed line~: phonon propagation length  $\lambda_{hydro}$ of eq.~\ref{eq:propagation_length_hydro}. 
Black dotted line~: phonon propagation length $\lambda_{gas}$ of eq.~\ref{eq:propagation_length_gas}.  
The red line segment marks the ranges of temperatures, from 3.0~K to 3.48~K, and sample dimension, 9.06~mm, in which a second sound peak has been reported in the binary direction ~\protect\cite{Narayanamurti:1972}. 
}
\end{figure}

The LTC, third-order anharmonic constants of the normal and Umklapp phonon interactions, 
and thermodynamical  averages have been calculated   
on a 28x28x28 \textbf{q}-point grid in the Brillouin zone, but for
the drift velocity below 2~K, which required a 40$\times$40$\times$40 grid.
Details of the calculation are given in the supplemental material. 
We have used the  wire geometry for
boundary scattering with Casimir's model, $\Gamma^b = \frac{ n^0_\nu (n^0_\nu+1) |\mathbf{c}_\nu^b| }{F \, L^\mathrm{Cas}}$, 
where $\mathbf{c}_\nu^b$ is the group-velocity in the direction of the smallest dimension. 
Specularity~\cite{Berman:1955,Rajabpour:2011,Bourgeois:2016} is neglected and 
$\frac{1}{F}$ accounts
for the geometrical ratio of $L^\mathrm{Cas}$ over the  finite (yet large) dimension along the heat transport direction~\cite{Sparavigna:2002,Fugallo:2013,Markov:2016,Markov:2016b,Markov:Note:2017:supplemental2}. 
Varying $\frac{1}{F}$ by 2 $\pm$ 1 (Fig.~\ref{fig:thermalk-bulk}) has little consequence on $\kappa_\perp$ above T=2~K.

Remarkably, our calculated LTC shows the same evolution as the experimental one 
over three orders of magnitude (Fig.~\ref{fig:thermalk-bulk}, resp. black dotted and green dashed lines),  
and  the various regimes  of heat transport are excellently described from ambient temperature down to 2~K. 
The LTC increases as $T^{-1}$ with the decrease of
temperature down to 10~K. Then in the absence of scattering other than phonon-phonon interaction, the LTC shows an exponential growth below 10~K
(black solid line). 
This behavior is directly due to the weakness 
of resistive (Umklapp) processes.

The account for boundary scattering makes the LTC value remain finite even in the asymptotic limit.
Moreover, the theoretical curves satisfactorily explain the experimental behavior of the LTC and in particular, the position of the
conductivity maximum, $T^{max}$, which is found to be 3.2~K for the 9.72 mm wire, in extremely satisfactory agreement with the maximum at 
3.6~K observed in experiment (Fig.~\ref{fig:thermalk-bulk}, resp. black dotted and green dashed lines). 
Further decrease of temperature leads to a decrease of the LTC with a decay law 
gradually approaching the  T$^{3}$ behavior expected for a regime in which boundary scattering dominates.

The first sign of the transition from the kinetic to hydrodynamic regime around 3~K in infinite samples is 
demonstrated in Fig.~\ref{fig:thermalk-bulk} by a large $(> 10^2)$ difference between our V- and SMA-BTE results for the
LTC 
(resp. black and red solid lines). This result shows that the repopulation of phonon states due to normal processes plays an important role, 
invalidating the SMA picture in which individual phonons have lifetimes and propagation lengths determined by all of the collisional processes  
(normal and Umklapp, eq.~\ref{eq:propagation_length_gas}). 
The same conclusion can be drawn by considering Fig.~\ref{fig:lw_avg} where, around 3~K, normal processes dominate 
over the resistive ones (Umklapp) by more than one order of magnitude, so that eq.~\ref{eq:condition_hydro} is fulfilled. 

The same difference in the LTC between V- and SMA-BTE 
is found in 
presence of sample boundaries (Fig.~\ref{fig:thermalk-bulk}, resp. black and red dotted lines) and, 
remarkably, 
the average extrinsic scattering rate $\Gamma^{b}_{av}$ calculated with Casimir's length  
$L^\mathrm{Cas}$~$=$~$3.86$~mm  (Fig.~\ref{fig:lw_avg}, black dotted line)  
lays in between the average 
normal $\Gamma^{n}_{av}$ and Umklapp $\Gamma^{U}_{av}$ scattering rates and thus, satisfy the criterion 
of eq.~\ref{eq:condition_Poiseuille}  for the existence of Poiseuille's flow and second sound observability~\cite{Guyer:1966}. 
The  temperature interval calculated with eq.~\ref{eq:condition_Poiseuille} is   1.5~K $<T<$ 3.6~K, 
in perfect agreement with the temperature range, 1.5~K$<T<$3.5~K, 
in which second sound has been observed experimentally in the trigonal direction (grey shaded region)~\cite{Narayanamurti:1972}.
For $L^\mathrm{Cas} = 9.06$~mm  in the binary direction (red dot-dashed line), the calculated interval is 
1.3~K $<T<$ 3.4~K, 
a temperature range slightly more extended than the experimental one, 
3.0~K$<T<$3.48~K~\cite{Narayanamurti:1972}.
In our calculations, Poiseuille's regime ends for temperatures lower than 1.5~K, where phonon scattering by sample boundaries 
becomes significant (Fig.~\ref{fig:lw_avg}).  

However, the average scattering rates discussed so far do not contain any information about repopulation mechanisms \cite{Markov:Note:2017:mean_free_path}~\nocite{Chiloyan:2016}. 
To account for them, we extract a heat wave propagation length $L_h$  that we define by the criterion~: 
\begin{equation}
\kappa(T,L^\mathrm{Cas}=L_h)=\kappa(T,\infty)/e, 
\label{eq:propagation_length_two}
\end{equation}
where $\kappa(T,\infty)$ denotes the LTC obtained for an infinite sample at a given temperature, and
$\kappa(T,L^\mathrm{Cas})$ denotes the LTC  obtained for a sample of finite dimension. The extracted HWPL 
$L_h$ is the cylindrical wire diameter 
$L^\mathrm{Cas}$ needed to reduce $\kappa(T,\infty)$ by $e$  
(Fig.~\ref{fig:regime_Lcas_T}, filled disks, and supplemental material for the trigonal direction~\cite{Markov:Note:2017:supplemental1}).

Remarkably, at low temperatures,  $L_h$ is found to be close to the phonon propagation length computed with Umklapp processes only (eq.~\ref{eq:propagation_length_hydro}).  These resistive processes
damp the heat wave, thus defining the wave traveling distance between the instant of heat wave generation to complete diffusion. A strong presence of normal processes, in turn, favors 
heat conduction and second sound behavior. With the increase of temperature, $L_h$ becomes close to the phonon propagation length accounting for both Umklapp and normal processes of eq.~\ref{eq:propagation_length_gas}, \textit{i.e.}
of an uncorrelated phonon gas (empty circles). 
We see that the behavior of $L_h$ as a function of temperature   
is the fingerprint of the transition from the hydrodynamic to kinetic regime. The temperature range and sample dimension in which observations of second sound are available in the binary direction 
(red line segment) are in extremely satisfactory agreement with the calculations, which support the occurrence of second sound at 3.0~K for a 9.72~mm wire. 
Fig.~\ref{fig:regime_Lcas_T} enables us also to predict the occurrence of second sound at other temperatures and sample sizes, for instance 
at 4.1~K in a 1~mm~size wire.         

\begin{figure}[t]
\begin{center}
  \subfigure[]{\includegraphics[width=0.5\textwidth]{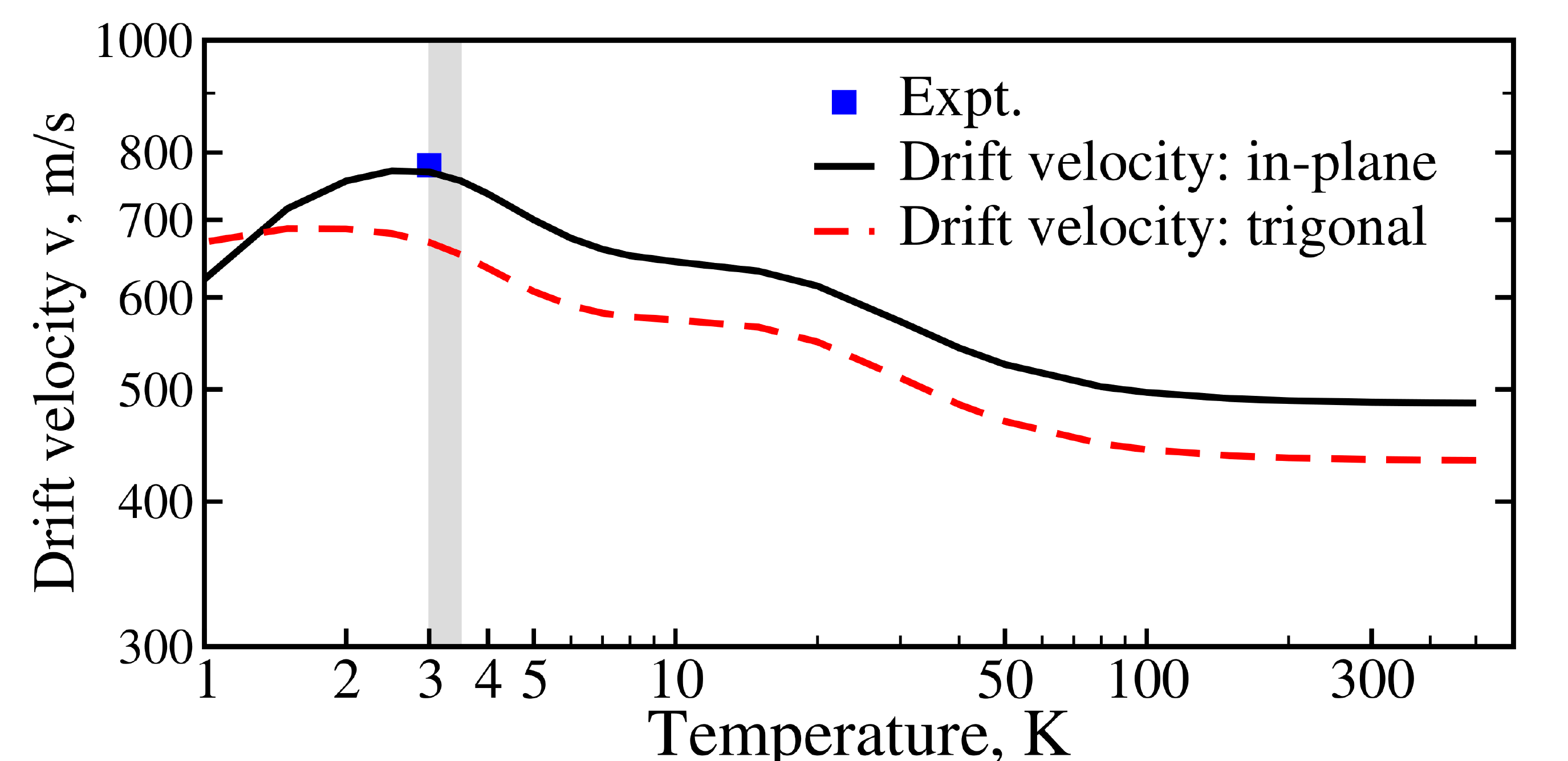}\label{fig:vss}}
  \subfigure[]{\includegraphics[width=0.5\textwidth]{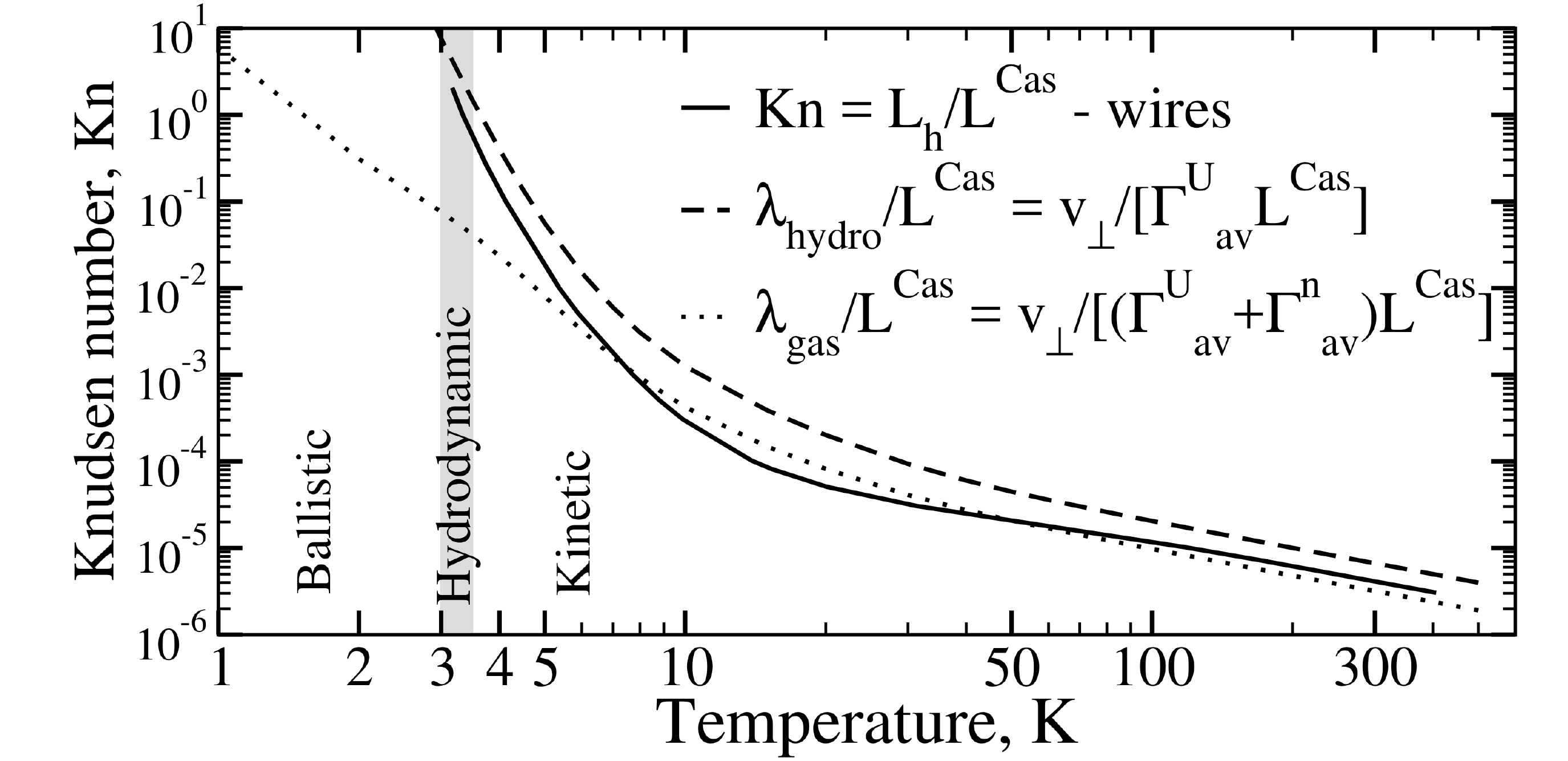}\label{fig:Kn_Mach}}
\caption{Heat flow characteristics in Bi as a function of the temperature. 
Panel~(a)~: drift velocity  $v$ in the binary  ($\perp$) and trigonal  ($\parallel$) directions 
(resp. black solid and red dashed lines).
Symbol~: saturated second-sound velocity measured at 3~K~\protect\cite{Narayanamurti:1972}.
Panel~(b)~: Knudsen number $L_h$/$L^\mathrm{Cas}$  for a wire of Casimir's length  
$L^\mathrm{Cas}$~$=$~$9.72$~mm (black solid line). The ratio of the
phonon propagation length in the hydrodynamic (resp. gas) regime over $L^\mathrm{Cas}$ is also given
(resp. black dashed and dotted lines).  
The shaded region 3.0~K $< T <$ 3.48~K corresponds to the temperature interval in which a second sound peak has been reported in the binary direction~\protect\cite{Narayanamurti:1972}.
}
\label{fig:heat_flow_characteristics}
\end{center}
\end{figure}

We emphasize that $L_h$ is a measurable quantity, provided that LTC can be measured in samples of many
different sizes, including very large ones. In that sense, the results presented in Fig.~\ref{fig:regime_Lcas_T} can be viewed as a Gedanken experiment in which~:
\textit{(i)} First, one need to determine the heat wave propagation length from the thermal conductivity measured in samples of different sizes, as described with
eq.~\ref{eq:propagation_length_two}. \textit{(ii)} Secondly, its combination with a measurement of the average phonon mean free path in a bulk sample,
given by eq.~\ref{eq:propagation_length_gas}, as done, for example, in attenuation measurement experiments \cite{Legrand:2016}, could, in principle, lead to the identification of the temperature and sample size ranges
in which Poiseuille's flow occurs.

We turn to the characterization of Poiseuille's flow, defined in the previous paragraph 
as the range of temperatures and propagation lengths where $L_h$ and $\lambda_{hydro}$ are close to each other. 
For this purpose we use common hydrodynamic quantities~: Knudsen number and drift velocity. The former 
is defined as the ratio between the HWPL  
and the characteristic dimension of transport,   
\begin{equation} 
Kn = \frac{L_{h}} {L^\mathrm{Cas}}. 
\label{eq:Knudsen_definition}
\end{equation}
Interestingly, the transition between the hydrodynamic and kinetic regime is found for a calculated Knudsen number $Kn$~$\approx$~$0.58$ at T~=~3.5~K  
in agreement with the criteria of phonon hydrodynamics $0.1 \lesssim Kn \lesssim 10$~\cite{Guo:2015} (Fig.~\ref{fig:heat_flow_characteristics}, bottom panel, black solid line). 
Our drift velocity calculated with eq.~\ref{eq:v} in the binary direction shows a maximum of $v_{\perp} = 770$ m/s at 3.0~K whose value matches well with the second sound 
velocity $v = 780$ m/s measured in Ref.~\onlinecite{Narayanamurti:1972}.  
At variance with the experiment~\cite{Narayanamurti:1972}, we find however a dependence on the propagation direction (top panel, black solid and red dashed lines). 

In conclusion, repopulation of phonon states by normal processes 
turns out to be particularly strong at low temperatures and leads to the occurrence of the hydrodynamic regime in bismuth. 
We have shown that this effect is remarkably well accounted for in the exact (variational) solution of the BTE.  
This enables us to extract from the lattice thermal conductivity a characteristic length, the heat wave propagation length,  whose behavior as a function of temperature, 
when compared to the phonon mean free path, is a fingerprint of the hydrodynamic to kinetic transition regime. We propose our method as a Gedanken experiment.
It provides an alternative to a standard heat
pulse propagation technique used in literature. Our calculated HWPL  matches with macroscopic sample dimensions
in which second sound was experimentally observed \cite{Narayanamurti:1972}. Finally, our calculated 
HWPL, Knudsen number and drift velocity allow to make the link with phonon hydrodynamics. 

We acknowledge discussions with A.~Cepellotti and A.~McGaughey. 
Support from the DGA  (France), from the Chaire \'Energie of the \'Ecole Polytechnique, from the program NEEDS-Mat\'eriaux (France) and from ANR-10-LABX-0039-PALM (project Femtonic) is gratefully acknowledged. 
Computer time was granted by \'Ecole Polytechnique through the LLR-LSI project and by GENCI (project No.~2210).

%\bibliography{/Users/vast/A_PHYSIQUE/biblio/mybiblio}
%\bibliography{/home/markov/Desktop/biblio/mybiblio}
%\bibliography{../../../SVN_Papier/biblio/mybiblio}
%\bibliography{/home/maxime/Desktop/WORK/polytechnique/biblio/mybiblio}

\begin{thebibliography}{39}
\bibitem[{Mar({\natexlab{a}})}]{Markov:Note:2015:Lcas_macro_one}
\bibinfo{note}{The longest dimension is along the binary direction. We used
  $L^{Cas} = 2 \sqrt{\frac{l_{1}l_{2}}{\pi}}$ for this rectangular wire, whose
  experimental cross-section is defined by the lengths $l_{1}$ and $l_{2}$.
  Choosing $L^{cas} = \sqrt{l_{1}l_{2}} = 8.7$mm has little effect on the scale
  of figure 1.}
\bibitem[{Mar({\natexlab{b}})}]{Markov:Note:2016:Lcas_second_sound_1}
\bibinfo{note}{The longest dimension is along the trigonal axis. For the
  cross-section, we have used a circular one oriented in (bisectrix, binary)
  plane with $L^{Cas}=d$, where $d=3.86$ mm is the experimental diameter of the
  cylindrical wire for sample~1. Choosing a spherical grain with $L^{Cas}=3.86$
  mm would yield a minor difference on the temperature interval in which
  hydrodynamic phonon transport occurs.}
\bibitem[{Mar({\natexlab{c}})}]{Markov:Note:2016:Lcas_second_sound_2}
\bibinfo{note}{The longest dimension is along the binary axis. For the
  cross-section, we have used a circular one oriented in (bisectrix, trigonal)
  plane with $L^{Cas}=d$, where $d = 9.06$ mm is the experimental diameter of
  the cylindrical wire for sample~2. Choosing a spherical grain with
  $L^{Cas}=9.06$ mm would yield a minor difference on the temperature interval
  in which hydrodynamic phonon transport occurs.}
\end{thebibliography}

\begin{widetext}
\clearpage

%\appendix
\section*{\Large Supplemental material}

We provide supplemental material to discuss convergence issues, the modeling of phonon-boundary scattering, and show that phonon repopulation by normal processes in bismuth at low temperatures
leads to the occurrence of the hydrodynamic regime also in the trigonal direction. 

\section{Details of the calculations}

The lattice thermal conductivity has been computed with the linearized Boltzmann transport equation and the variational method 
(VAR-BTE) on a 28x28x28 \textbf{q}-point grid in the BZ with a Gaussian
broadening of the detailed balance condition taken
to be $\sigma$~=1~cm$^{1}$~\cite{Fugallo:2013}. Details of the calculation have been reported in Refs.~\onlinecite{Markov:2016,Markov:2016b}.
Third-order anharmonic constants of the normal and Umklapp phonon interactions
have been
computed on a 4$\times$4$\times$4 \textbf{q}-point grid in the BZ. 
The 4x4x4 grid amounts to 95 irreducible
($\mathbf{q}_1$,$\mathbf{q}_2$,$\mathbf{q}_3$)
phonon-triplets~\protect\cite{Paulatto:2013}, where $\mathbf{q}_i$, i=1,3 are
phonon wavevectors, and with $\mathbf{q}_1=\mathbf{q}_2 \pm \mathbf{q}_3 +
\mathbf{G}$. $ \mathbf{G}$ is a vector of the reciprocal lattice. The
third-order anharmonic constants were Fourier-interpolated on the
28$\times$28$\times$28 denser grid necessary for converged integrations in
$\Gamma$. The convergence of $\Gamma^U$, $\Gamma^n$, $\lambda_{gas}$,
$\lambda_{hydro}$ and of $\kappa$ computed within VAR-BTE has been checked at
T = 2~K on a 34$\times$34$\times$34 grid. The thermodynamic averages were
calculated on the grid 28x28x28 and the convergence was checked on the
34x34x34 grid.
Below 2~K however, 
the drift velocity of Fig. 4(a) of the main text required a 40$\times$40$\times$40 grid.

\section{Lattice thermal conductivity  in the trigonal direction}

\begin{suppfigure}[h]
\begin{center}
 \includegraphics[width=0.5\textwidth]{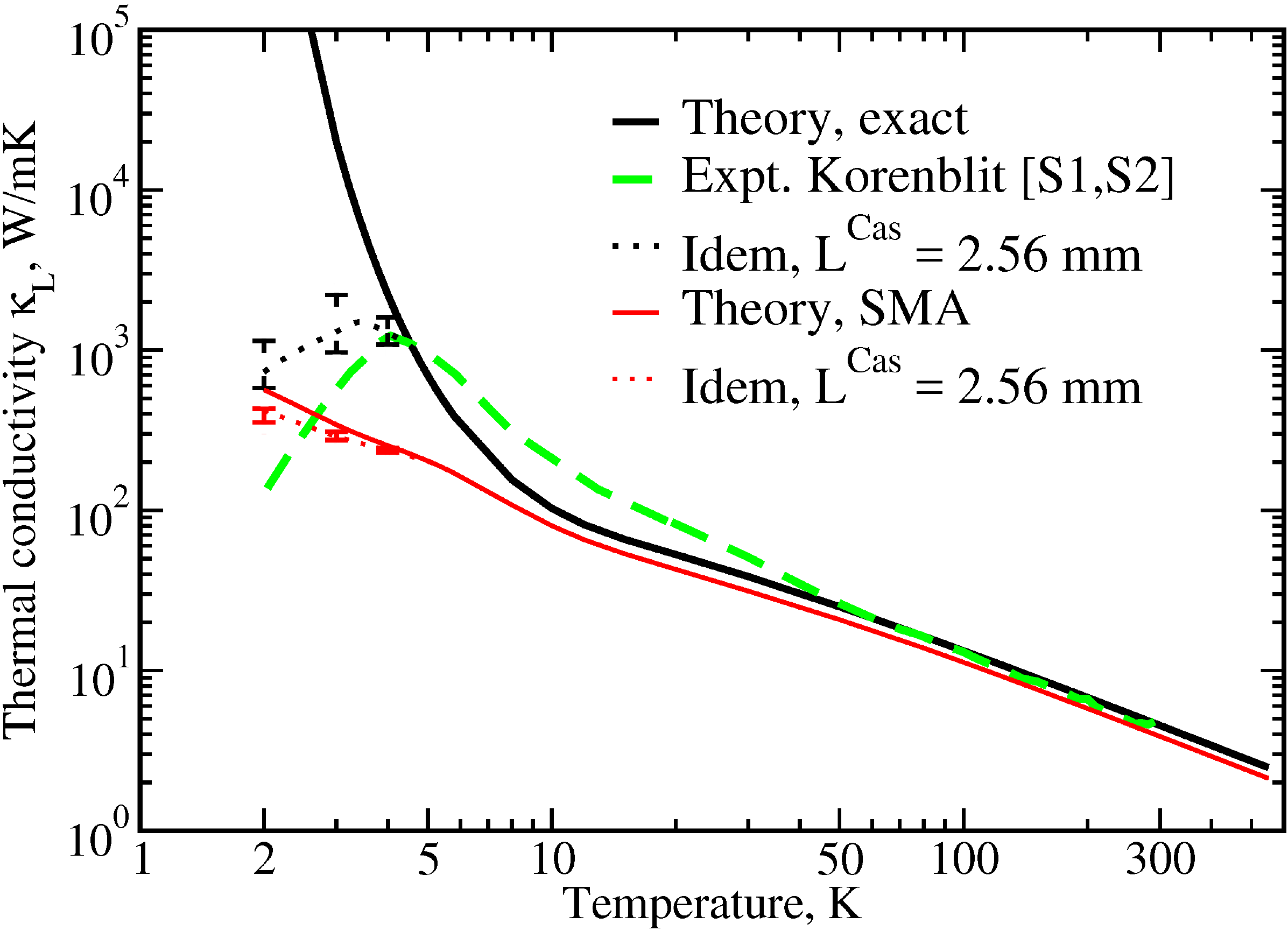}
\caption{\label{fig:kappa_trigonal}Temperature dependence of the lattice thermal conductivity (LTC) in the
trigonal direction for a single crystal without (solid lines) or with (dotted lines) millimeter-sized sample boundaries (MSSB).   
Black curves~: exact variational calculation (V-BTE). Red curves~: single mode approximation (SMA-BTE).  
MSSB modeled with the wire geometry and $L^\mathrm{Cas}$~$=$~2.56~mm.
Green dashed lines~:  LTC extracted by us from expt. of Ref.~\protect\onlinecite{Collaudin:2014} for $T>10$~K;
for $T<10$~K, LTC from a sample  having a circular cross-section $2.56$ mm$^{2}$ (Ref.~\protect\onlinecite{Korenblit:1970}).
We used the T-independent bulk value of 3~W(K.m)$^{-1}$~\protect\cite{Markov:2016,Markov:2016b} of the electronic contribution 
to extract the LTC from the total thermal conductivity of Bi~\protect\cite{Collaudin:2014}. Error bars in our calculations represent 
variation of the geometrical factor $\frac{1}{F}$ from 1 to 3, \textit{i.e.} $\frac{1}{F}$=2$\pm$1, in Casimir's model. }
\end{center}
\end{suppfigure}

\begin{suppfigure}[h]
\begin{center}
 \includegraphics[width=0.5\textwidth]{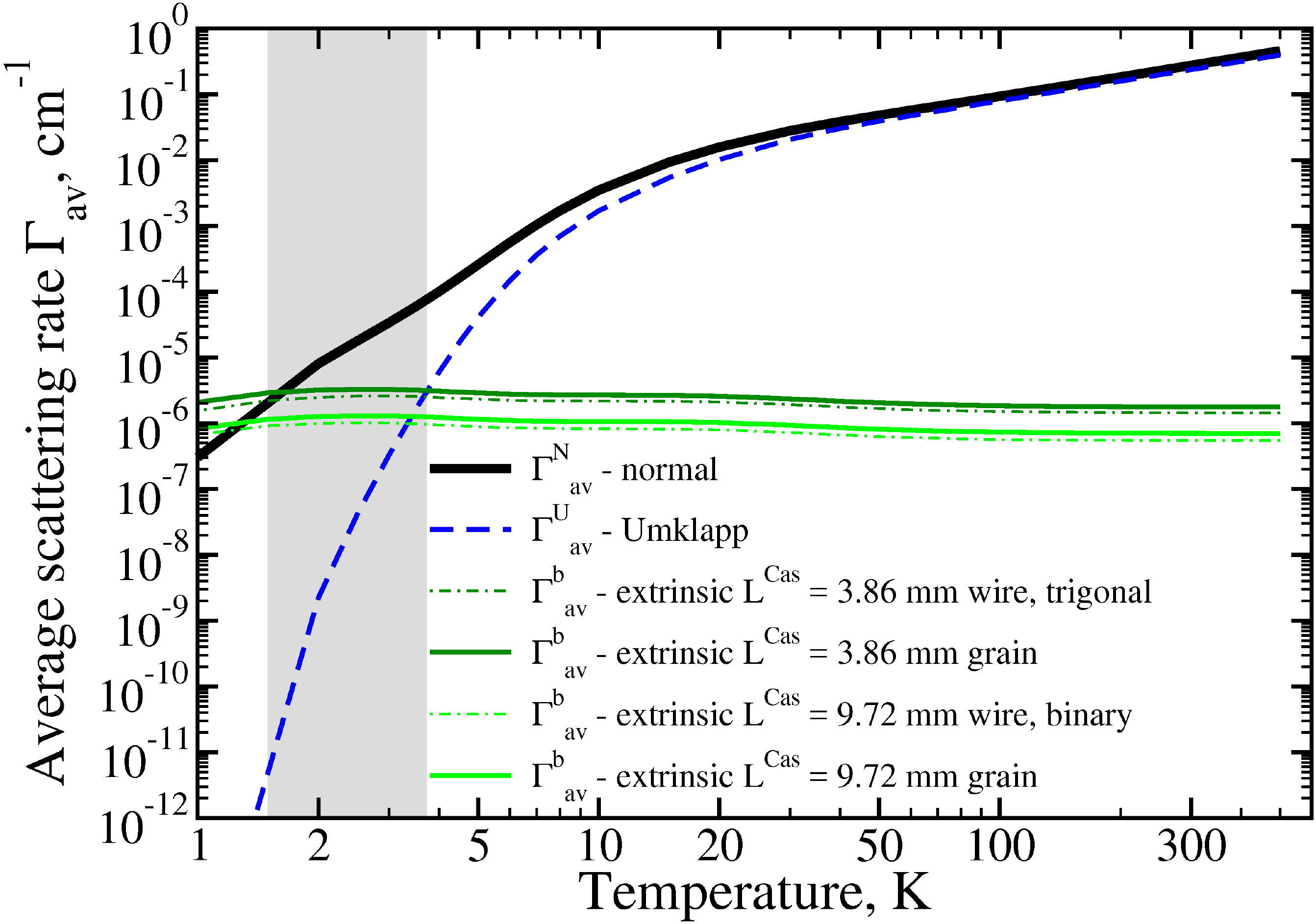}
\caption{\label{fig:lw_avg_geometry}Temperature dependence of the thermodynamic average of the anharmonic scattering rates for  normal and Umklapp processes
(resp. black solid and blue dashed line) and of the (boundary) extrinsic scattering rates (ESR) (green solid and dot-dashed lines). ESR have been calculated for a wire
(green dotted-dashed lines) and a spherical grain (green solid lines) geometry with effective sizes $L^\mathrm{Cas}$~$=$~$3.86$ mm (dark green) and $L^\mathrm{Cas}$~$=$~$9.72$~mm
(light green). Wires are oriented in the trigonal and binary directions respectively. The difference between two geometries is small.
The shaded region corresponds to the temperature interval in which a second sound peak has been reported,  1.5~K $< T <$ 3.5~K  for a sample of length
3.86~mm in the trigonal propagation direction~\cite{Narayanamurti:1972}.
}
\end{center}
\end{suppfigure}

\begin{suppfigure}[h]
\begin{center}
 \includegraphics[width=0.5\textwidth]{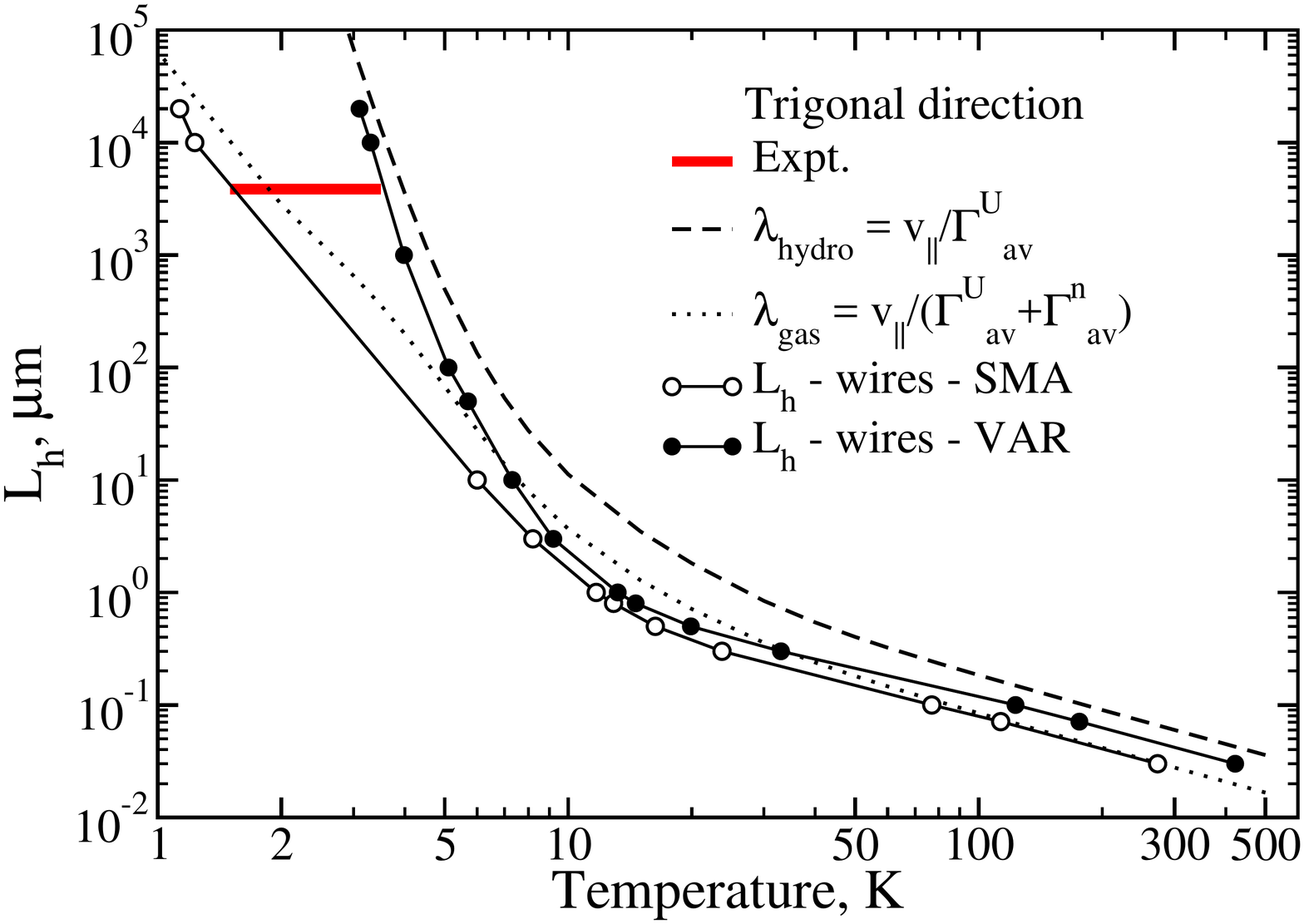}
\caption{\label{fig:regime_Lcas_T_trigonal_dir}
Heat wave propagation length $L_h$ extracted from the LTC calculations in the trigonal direction
as a function of temperature.
Solid line with black filled disks~:
$L_h$ obtained with V-BTE, accounting for phonon repopulation. Solid line with empty circles :
$L_h$ obtained with SMA-BTE.
Black dashed line~: phonon propagation length  $\lambda_{hydro}$.
Black dotted line~: phonon propagation length $\lambda_{gas}$.
The red line segment marks the ranges of temperatures, from 1.5~K to 3.5~K, and sample dimension, 3.86~mm, in which a second sound peak has been reported in the trigonal direction ~\protect\cite{Narayanamurti:1972}.
}
\end{center}
\end{suppfigure}

Fig.~S~\ref{fig:kappa_trigonal} is equivalent to Fig.~1 of the main text. It shows the lattice thermal conductivity in the trigonal direction calculated using the exact solution of the 
BTE (black curves), the single mode approximation (red curves) and experimental lattice thermal conductivity obtained from the measured total conductivity from Ref.~\onlinecite{Collaudin:2014} 
with the subtracted T-independent bulk value of 3~W(K.m)$^{-1}$~\protect\cite{Markov:2016,Markov:2016b} of the electronic contribution (green dashed curve). At low temperatures, T$<$ 10 K,
data were extracted from Ref.~\onlinecite{Korenblit:1970} for the cylindrical wire with diameter d = 2.56 mm. Solid lines represent the calculations with phonon-phonon scattering only. Dotted
lines represent the calculations accounting for the phonon scattering by boundaries in addition to the phonon-phonon scattering. We use $L^{Cas} = 2.56$~mm and the geometrical $\frac{1}{F}$ = 2.
Varying $\frac{1}{F}$  from 1 to 3, we introduce an error bar in our calculations.

\section{Sample geometry}

In our study, for the sake of unity, we present only the results for the wire geometry in all of the figures. 
The wire geometry corresponds to the one in which the thermal conductivity has been measured (Refs. 11 and 22).
Phonon-boundary scattering is thus defined by the shortest dimension of the sample $L^{Cas}$ that is assumed to be
perpendicular to the transport direction, and the wire length $l$ is assumed to be much larger than
$L^{Cas}$ in the calculations. However, the conclusions based on the results presented in Fig.~2, about the occurrence
of the hydrodynamic regime, are exactly the same with the grain geometry, \textit{i.e.} if the sample length in the heat propagation direction 
is taken to be $l=L^{Cas}$. Indeed, in  Fig.~S~\ref{fig:lw_avg_geometry}, the lines for grains and for nanowires (green solid lines and green dotted-dashed lines) 
are almost indistinguishable. 

In samples in which second sound has been measured (Ref. \onlinecite{Narayanamurti:1972}), the samples were cut in the heat transport direction.
The finite length along the transport direction in real samples determined whether
the heat pulse can be detected or not. If the propagation length is smaller than the wire length, the heat pulse reaching the sample edge is already
damped and, thus, can not be registered by a receiver. While in the opposite case, the temperature wave is still observable and can be detected. 
In Fig. 3 of the main text, we evaluate the propagation length of the temperature wave \textit{i.e.} the distance at which the amplitude of wave is decreased by a factor $\frac{1}{e}$.  
When  $\lambda>l$, the second sound is not dumped at length $l$ and,thus, can be registered by the receiver. 

\section{Hydrodynamic regime in the trigonal direction}

Fig.~S~\ref{fig:regime_Lcas_T_trigonal_dir} is
equivalent to Fig.~3 of the main text, with the heat wave propagation length (HWPL) $L_h$ computed from the lattice thermal conductivity (LTC)
in the trigonal direction with eq.~7 of the main text.
At low temperatures the HWPL $L_h$ is found to be close to the phonon propagation length, $\lambda_{hydro}$,  computed with Umklapp processes only, while
with the increase of temperature, $L_h$ becomes close to the phonon propagation length, $\lambda_{gas}$, accounting for both Umklapp and normal processes, \textit{i.e.}
of an uncorrelated phonon gas (empty circles). Latter quantities $\lambda_{hydro}$ and $\lambda_{gas}$ are computed resp. with eqs.~4~and~5~of the main text.

\section{Surface roughness and specularity}

In Figs.~1~(main text) and~S~\ref{fig:kappa_trigonal}, we used Casimir's model with respectively $L^{Cas} = 9.72$~mm or  $L^{Cas} = 2.56$~mm and the geometrical factor $\frac{1}{F}$ = 2.
The Casimir model has been extensively and successfully employed with these two parameters 
for the description of phonon-boundary scattering in a wide variety of materials (see Refs.~\onlinecite{Sparavigna:2002,Fugallo:2013} for diamond; Ref.~\onlinecite{Park:2014} for silicon; Ref.~\onlinecite{Sparavigna:2002b} for silicon carbide),
including our recent calculation for polycrystalline thin films of bismuth~\cite{Markov:2016}. When larger than the unity,  $\frac{1}{F}$ accounts 
for the geometrical ratio of $L^\mathrm{Cas}$ over the  finite (yet large) dimension along the heat transport direction~\cite{Sparavigna:2002,Fugallo:2013,Markov:2016,Markov:2016b} and also, to a smaller extent, to  a (small) specularity of 
an otherwise almost completely diffusive (rough) surface~\cite{Berman:1955,Rajabpour:2011,Bourgeois:2016}.
To demonstrate that our prediction does not depend on the value of the geometrical factor we change $\frac{1}{F}$ from 1 to 3, increasing and
decreasing the role of boundary scattering correspondingly. We set these values as an error bar for our calculations. \\

\end{widetext}

\end{document}